\documentclass[a4paper,12pt]{article}

\usepackage[margin=1in]{geometry} 
\usepackage{amsmath,amsthm,amssymb,setspace,scrextend,makeidx,graphicx,longtable,float,hyperref,tikz,xcolor}

\definecolor{lime}{HTML}{A6CE39}
\DeclareRobustCommand{\orcidicon}{
	\begin{tikzpicture}
	\draw[lime, fill=lime] (0,0) 
	circle [radius=0.16] 
	node[white] {{\fontfamily{qag}\selectfont \tiny ID}};
	\draw[white, fill=white] (-0.0625,0.095) 
	circle [radius=0.007];
	\end{tikzpicture}
	\hspace{-2mm}
}

\usepackage[title]{appendix}
\usepackage[section]{placeins}
\usepackage{multirow}

\usepackage{tabularx}
\newcolumntype{L}[1]{>{\raggedright\arraybackslash}p{#1}}
\newcolumntype{C}[1]{>{\centering\arraybackslash}p{#1}}
\newcolumntype{R}[1]{>{\raggedleft\arraybackslash}p{#1}}
\usepackage{dirtytalk}

\usepackage[numbers]{natbib}
\hypersetup{pdfusetitle,
            pdfauthor={Johannah Sprinz},
            pdflang={en},
            pdfsubject={Computational Ethics},
            pdfkeywords={microaggressions, computational ethics, simulated societies}
}

\begin{document}

% ------------------------------------------ %
%                 START HERE                  %
% ------------------------------------------ %

\title{\huge{"Y'all are just too sensitive":}\\
\Large{A computational ethics approach to understanding how prejudice against marginalized communities becomes epistemic belief}}
\date{June 30, 2022}
\author{
    \href{https://orcid.org/0000-0002-6425-7054}{Johannah Sprinz\orcidicon}
}
\maketitle

\begin{center}

\vspace{-1cm}
LMU Munich\footnote{This research was originally conducted as part of the Master's practical seminar "Computational Ethics" at LMU Munich in the winter term of 2021, supervised by Prof. Dr. François Bry.
Released as an open-access preprint licensed \href{https://creativecommons.org/licenses/by/4.0/}{CC BY 4.0}; (c) 2022
% \href{https://de.linkedin.com/in/matthias-fruth-825711180}{Matthias Fruth} and 
\href{https://spri.nz}{Johannah Sprinz}.
Significant contributions by \href{https://de.linkedin.com/in/matthias-fruth-825711180}{Matthias Fruth} are acknowledged.
}

% \footnotesize
% Submitted to the \href{https://beta.briefideas.org/}{\scshape Journal of Brief Ideas}\\
% \textbf{doi}: \href{https://doi.org/PENDING}{PENDING}\\
\end{center}
\vspace{.5cm}

\begin{addmargin}[6em]{6em}
    \item
    \small
    \paragraph{Abstract}
    Members of marginalized communities are often accused of being “too sensitive” when subjected to supposedly harmless acts of microaggression. This paper explores a simulated society consisting of marginalized and non-marginalized agents who interact and may, based on their individually held convictions, commit acts of microaggressions. Agents witnessing a microaggression might condone, ignore or condemn such microaggressions, thus potentially influencing a perpetrator’s conviction. A prototype model has been implemented in NetLogo, and possible applications are briefly discussed.
    \normalsize
\end{addmargin}

\vfill

\pagebreak

\section{Introduction}

\paragraph{Motivation}

Members of marginalized communities are often accused of being \textit{too sensitive} when it comes to supposedly harmless acts of microaggression. Sexist comments are regarded as \textit{locker room talk}. Racist jokes are viewed as \textit{harmless fun}. Misgendering of trans people is downplayed as \textit{a mistake that can happen}.

Microaggressions have been shown to cause significant emotional harm. Nonetheless, victims are often met with incomprehension, defensiveness, or even ridicule when pointing out microaggressions. Stereotypes like the \textit{snowflake liberal} and the \textit{shrill feminist who can't take a joke} paint members of marginalized communities as irrational and dismiss the pain and harm inflicted upon them.

\paragraph{Contribution}

This paper applies computer simulation to the phenomenon of microaggressions. Observing interactions of marginalized and non-marginalized agents in simulated societies with varying prevailing convictions should enable a better understanding of how prejudice against marginalized communities becomes an epistemic belief.

\paragraph{Outline}

The following section highlights notable related research in the fields of computational ethics and the social sciences. Next, the methodological principles for a model are outlined. A prototype for a NetLogo simulation has been implemented and it's behavior is described. Finally, possible directions for future work on this topic are discussed.

\section{Related Work}

\paragraph{Computational Ethics}

Computational ethics makes use of "agent-based simulation [to apply a] a computational perspective to ethics theory" \cite[76]{quigley_computational_2007} by simulating agents capable of adopting malleable ethical principles and observing how interactions impact their ethical principles. Such simulations can provide descriptive\footnote{It can only be concluded how the input relates to the output (descriptive), not what input is required to achieve a certain output (perscriptive) \cite[76]{quigley_computational_2007}.} insights into how individual ethical principles impact societal dynamics.

\begin{figure}[!htb]
    \centering
    \includegraphics[width=\textwidth]{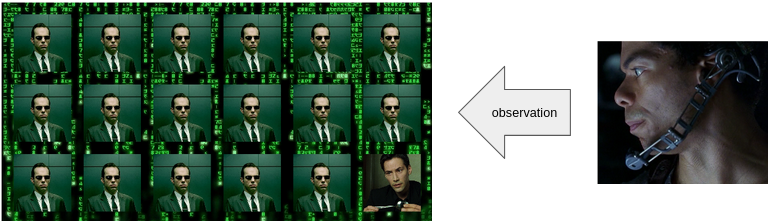}
    \caption{Computational ethics observes simulated societies consisting of interacting agents with individual ethical principles. Photos (c) 1999 Warner Bros.}
    \label{fig:simulated_society}
\end{figure}

\paragraph{Microaggressions}

Microaggressions were initially conceptualized to describe unacknowledged racism experienced by African Americans \cite{pierce_offensive_1970}. Since other marginalized groups are subjected to similar types of discrimination, the definition has been extended and generalized over the years. More recent definitions literature defines microaggressions as "brief and commonplace daily verbal, behavioral, and environmental indignities, whether intentional or unintentional, that communicate hostile, derogatory, or negative racial, gender, sexual orientation, and religious slights and insults to the target person or group" \cite{sue_microaggressions_2010}. Since prevailing biases influence all members of society, even well-intentioned people will unconsciously commit microaggressions. The relentlessness of these unintentional indignities can cause emotional harm comparable to more obvious intentional and outright malicious types of discrimination such as slurs and acts of physical violence \cite{sue_microaggressions_2010} \cite{kapusta_misgendering_2016} \cite{kelly_2021}.

\paragraph{Colorblind Ideology}

Some people, however, profess the idea of a \textit{colorblind ideology} by which they claim to be uninfluenced by societal bias or even deny the existence thereof\cite{eisen_combating_2020}. This kind of ideology provides individuals with numerous ways to downplay and and deny the impact of racism \cite{eisen_combating_2020}. Eisen identifies four types of colorblind racism: abstract liberalism, naturalization, cultural racism, and minimization of racism \cite{eisen_combating_2020}. "By employing these frames, individuals divert their attention away from racism and effectively engage in racialized discourse without appearing racist" \cite{eisen_combating_2020}. 
Individuals often react defensively when their beliefs are challenged. As such, reluctance to acknowledge microaggressions is a common reaction \cite{eisen_combating_2020}. Such defensive reactions might include the argument that those affected by microaggressions are just too sensitive. However, people who proffer this argument rarely understand the nature of microaggressions and how constant exposure may affect people. For an unintentional perpetrator being criticized, it might seem that the person is overreacting because they are usually oblivious to the countless other incidents. Closer examination however reveals that "victims of microaggression need not to just develop ‘thick skin’ to overcome the epistemic harm of microaggression but need some form of support or epistemic resources to overcome these disadvantages" \cite[21]{kelly_2021}.% Eisen \cite{eisen_combating_2020} described an experiment for classrooms to teach about the nature of microaggressions. Marbles are constantly dropped on a stretched foil over a glass. While a single marble is not able to tear the foil, constant dropping of marbles will eventually lead to a hole and finally all marbles will make their way into the glass.

%the "[...] constant and continuing everyday reality of slights, insults, invalidations, and indignities visited upon marginalized groups by well-intentioned, moral, and decent family members, friends, neighbors, coworkers, students, teachers, clerks, waiters and waitresses, employers, health care professionals, and educators."\cite{sue_microaggressions_2010}

\paragraph{Approach}

By a qualitative analysis, one can understand how the microaggressions and prejudices against people pointing them out may emerge on the individual level. Applying a computational ethics simulation might provide further insights into this dynamic on the societal level.

\section{Methodology}

This chapter introduces the conceptual ideas that constitute the model and describes the prototype implementation in the multi-agent simulation framework NetLogo \cite{wilensky_netlogo_1999}.

\subsection{Model}

The model is constructed according to the following principles:

\begin{enumerate}
  \item A society consists of a number of agents with individually held believes. We observe how much agents agree with the following two statements:
  \begin{enumerate}
      \item[1.] "Microaggressions do not constitute a wrong."
      \item[2.] "Marginalized agents are overly sensitive."
  \end{enumerate}
  \item An agent might identify as part of the marginalized community within the society.
  \item Agents interact with each other randomly.
  \item A microaggression does not need to be targeted towards an individual agent.
  \item The decision to commit a microaggression does not have to be conscious, nor does the perpetrator have to be aware that their action constituted a microaggression. Whether or not an agent will commit a microaggression depends on their conviction that microaggressions do not constitute a moral wrong. As such, they may commit a microaggression if and only if their agreement with statement 1 is sufficiently high.
  \item An Agent witnessing a microaggression will react positively, neutrally, or negatively based on whether or not they consider microaggressions a moral wrong, i.e., their agreement with statement 1. A perpetrator's conviction may be influenced by the reaction they received, and the reacting agent's convictions might themselves be influenced by the reaction they have given. Negative feedback might be accepted or rejected by the perpetrator. If a perpetrator rejects negative feedback from a marginalized agent, the conviction that marginalized agents are overly sensitive may be reinforced.
  \item Consumed media and other background noise might subconsciously influence agents' convictions positively or negatively independent from interactions.
  \item For simplicity, only one type of marginalization is observed at a time. Intersectionality\footnote{The concept of intersectionality plays an important role when discussing topics of marginalization, as it enables the analysis of situations where a person is affected by multiple types of marginalization at once. The term was coined by Kimberle Crenshaw in 1989, highlighting that the marginalization experienced by black women "cannot be understood through an analysis of patriarchy rooted in white experience" \cite[156f]{bartlett_demarginalizing_2018}.} is considered out of scope as of now.
\end{enumerate}

\subsection{Simulation} \label{text:simulation}

This section explains the prototype implementation\footnote{The model is available under \url{https://neothethird.gitlab.io/ceth-seminar/model.nlogo}.} of the previously described model in as an agent-based simulation in NetLogo.

% \begin{figure}[!htb]
%     \centering
%     \includegraphics[width=\textwidth]{img/board.png}
%     \caption{Screenshot of the NetLogo model's visualized society, the output textbox, and some important control sliders.}
%     \label{fig:board}
% \end{figure}

\paragraph{Fundamental Principles}

Marginalized and Non-Marginalized agents can be initialized with normal-distributed agreement percentages for two convictions:

\begin{itemize}
    \item \texttt{c1}: "microaggressions do not constitute a wrong"
    \item \texttt{c2}: "members of the marginalized group are overly sensitive"
\end{itemize}

The simulation runs until one of three possible end conditions is reached:
\begin{enumerate}
    \item \texttt{equilibrium: no potential perpetrators}: Microaggressions have effectively been eradicated; there are no agents with \texttt{c1 >= action\_threshold}.
    \item \texttt{equilibrium: no negative reactors}: Microaggressions have become so normalized that no-one speaks up against them; there are no agents with \texttt{c1 <= negative\_threshold}.
    \item \texttt{deadlock: society is too polarized for change}: All agents either have very strong agreement or disagreement with \texttt{c1}. Change is improbable due to the high polarization.
\end{enumerate}

Agents can interact with one another one-to-one and may commit microaggressions in these interactions. An agent witnessing a microaggression may react positively, negatively, or neutrally. This behavior is configured through thresholds that control Poisson-distributed probability values calculated for every interaction. Example: Suppose the \texttt{action\_threshold} is set to \texttt{75\%}. All agents with \texttt{c1 >= 75} may now theoretically become perpetrators. Whether or not an agent actually commits a microaggression is determined by \texttt{c1 >= individual\_action\_threshold}, where \texttt{individual\_action\_threshold} is a poisson-distributed probability variable with a mean of \texttt{(action\_threshold + (100 - action\_threshold) / 2) = 75 + 12.5 = 87.5}. Thus, the higher the \texttt{c1} of an agent, the higher their chance of becoming a perpetrator.

Reactions are Poisson-distributed as well. Positive reactions are controlled by the \texttt{individual\_positive\_threshold}, a Poisson-distributed probability variable with a mean of \texttt{(positive\_threshold + (100 - positive\_threshold) / 2)}; halfway between \texttt{100} and \texttt{positive\_threshold}. Negative reactions are controlled by \texttt{individual\_negative\_\\threshold}, a poisson-distributed probability variable with a mean halfway between \texttt{0} and \texttt{negative\_threshold}, i.e., \texttt{negative\_threshold / 2}.

\paragraph{Visualization}

\begin{figure}[!htb]
    \centering
    \includegraphics[scale=.42]{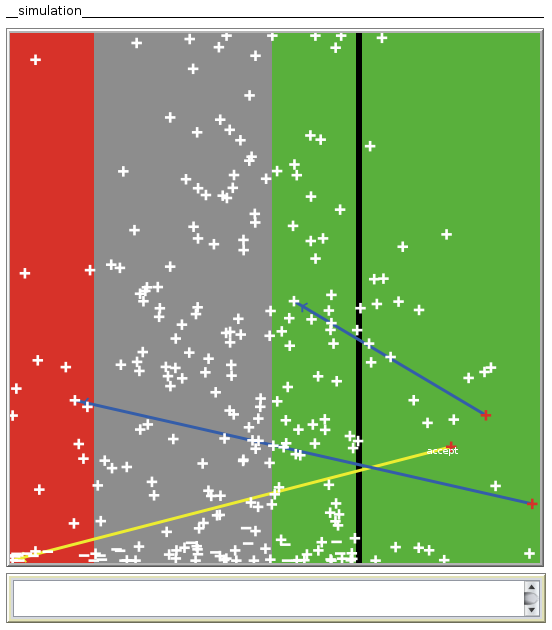}
    \caption{Screenshot of the NetLogo model's visualized society.}
    \label{fig:simulation}
\end{figure}

Agents are rendered on a a two-dimensional plane (see Figure \ref{fig:simulation}) based on their convictions with the \texttt{x}-axis representing \texttt{c1} and the \texttt{y}-axis representing \texttt{c2}.

\begin{figure}[!htb]
    \centering
    \includegraphics[scale=.45]{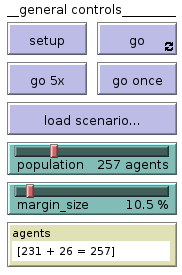}
    \caption{Screenshot of the NetLogo model's general controls. The society consists of 257 agents, 10.5\% (=26) of which are marginalized, leaving 231 non-marginalized agents.}
    \label{fig:general}
\end{figure}

\begin{figure}[!htb]
    \centering
    \includegraphics[scale=.66]{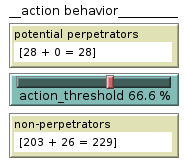}
    \caption{Screenshot of the NetLogo model's action controls.}
    \label{fig:action}
\end{figure}

\begin{figure}[!htb]
    \centering
    \includegraphics[scale=.66]{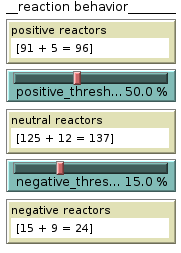}
    \caption{Screenshot of the NetLogo model's reaction controls.}
    \label{fig:reaction}
\end{figure}

The \texttt{action\_threshold} is represented by a vertical black line. Agents right of this line are potential perpetrators. The green area represents the \texttt{positive\_threshold}; Agents in this area might react positively to microaggressions. The red area represents the \texttt{negative\_threshold}; Agents in this area might react negatively to microaggressions. The grey area contains agents that will always react neutrally to microaggressions, as their \texttt{c1} is below the \texttt{positive\_threshold} and above the \texttt{negative\_threshold}. Marginalized agents are represented by a \textbf{-}, non-marginalized agents by a \textbf{+}. If an agent is tinted red, it will commit a microaggression in this tick. Otherwise, agents are tinted white. An arrow points from the perpetrator of a microaggression to the reacting agent. A pink arrow indicates a positive reaction to the microaggression; blue indicates neutral, and yellow indicates negative. If a perpetrator receives a negative reaction and accepts the criticism, they will be marked with the label "accept". Otherwise, with the label "reject". The output field beneath the map displays end conditions for the simulation. Monitors reporting values in the form of\texttt{[non\_marginalized + marginalized = total]} (see Figures \ref{fig:general}, \ref{fig:action}, and \ref{fig:reaction}) are used to show how many agents a specific claim applies to at any given time; overall and broken down by whether or not they are marginalized. The user interface also includes various plots (see Section \ref{text:validation}) that are updated in real time.

\vspace{-.25cm}
\paragraph{General Controls}

Sliders \texttt{population} and \texttt{margin\_size} (see Figure \ref{fig:general}) control the number of agents and the percentage of marginalized agents. The \texttt{setup} button will initialize a society based on the current configuration, \texttt{go} will start the simulation. the \texttt{go once} and \texttt{go 5x} buttons can be used for step-by-step execution. Pre-configured scenarios can be loaded using the \texttt{load scenario...} button. Every scenario is explained after selection.

\vspace{-.75cm}

\paragraph{Conviction Initialization}

\begin{figure}[H]
    \centering
    \includegraphics[scale=.45]{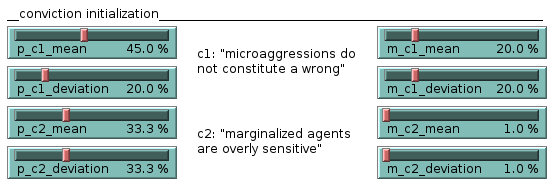}
    \caption{Screenshot of the NetLogo model's conviction sliders.}
    \label{fig:conviction_init}
\end{figure}

Convictions one (\texttt{c1}) and two (\texttt{c2}) for marginalized (\texttt{m}) and non-marginalized (\texttt{p}) agents are initialized as normal-distributed values with a \texttt{mean} and a \texttt{deviation} using sliders (see Figure \ref{fig:conviction_init}) in the form of \texttt{<p|m>\_<c1|c2>\_<deviation|mean>}.

\paragraph{Action Behavior}

The \texttt{action\_threshold} slider (see Figure \ref{fig:action}) controls the minimum required \texttt{c1} for an agent to commit a microaggression.

\paragraph{Reaction Behavior}

The \texttt{positive\_threshold} (see Figure \ref{fig:reaction}) controls the minimum required \texttt{c1} for an agent to react positively to a microaggression. Similarly, the \texttt{negative\_\\threshold} specifies the maximum allowed \texttt{c1} for an agent to react negatively to a microaggression.

\paragraph{Miscellaneous Controls}

\begin{figure}[!htb]
    \centering
    \includegraphics[scale=.66]{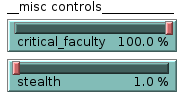}
    \caption{Screenshot of the NetLogo model's miscellaneous controls.}
    \label{fig:misc}
\end{figure}

The \texttt{critical\_faculty} slider (see Figure \ref{fig:misc}) allows setting the likelihood of a perpetrator accepting criticism when made aware of a microaggression they committed. The \texttt{stealth} slider enables the simulation of types of marginalization that are less recognizable to other agents by specifying the likelihood of an agent not recognizing another agent as marginalized.

\paragraph{Reactions}

\begin{figure}[!htb]
    \centering
    \includegraphics[width=\textwidth]{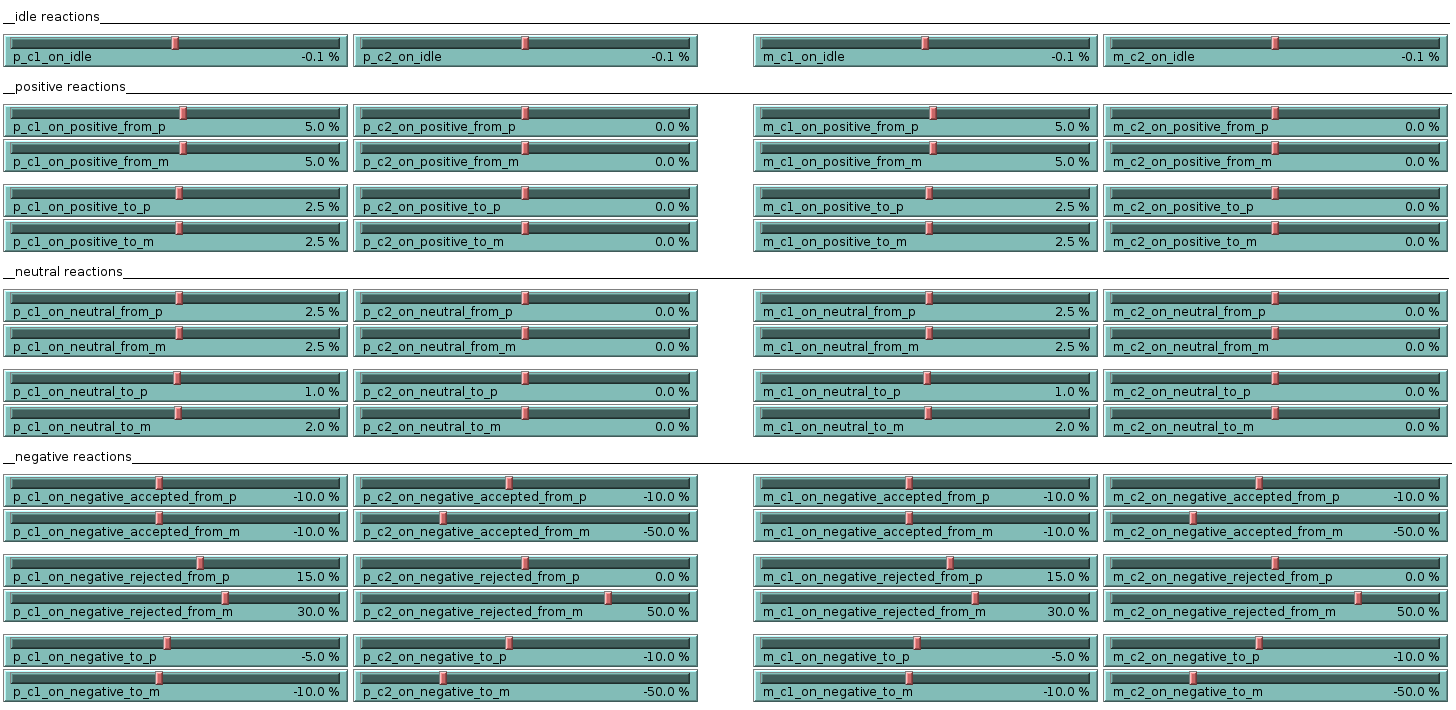}
    \caption{Screenshot of the NetLogo model's reactions sliders.}
    \label{fig:reactions}
\end{figure}

Changes in convictions one (\texttt{c1}) and two (\texttt{c2}) due to interactions can be set individually for marginalized (\texttt{m}) and non-marginalized (\texttt{p}) agents. If no microaggression occurs in an interaction, the \texttt{idle} value is applied. Otherwise, a \texttt{positive}, \texttt{neutral}, or \texttt{negative} reaction might be given \texttt{to} or received \texttt{from} a marginalized (\texttt{m}) or non-marginalized (\texttt{p}) agent. A received \texttt{negative} reaction might be \texttt{accepted} or \texttt{rejected}.
Sliders (see Figure \ref{fig:reactions}) take the form \texttt{<p|m>\_<c1|c2>\_on\_<idle|<<positive\_<to|from>|\\neutral\_<to|from>|negative\_<to|accepted\_from|rejected\_from>>\_<p|m>>}.

Example: A privileged agent, Bob, commits a microaggression in an interaction with a marginalized agent, Alice, who reacts neutrally. Bob's convictions change based on \texttt{p\_c1\_on\_neutral\_from\_m} and \texttt{p\_c2\_on\_neutral\_from\_m} respectively, while Alice's convictions change based on \texttt{m\_c1\_on\_neutral\_to\_p} and \texttt{m\_c2\_on\_neutral\_to\_p}.

\paragraph{Noise}

Background noise changes to convictions one (\texttt{c1}) and two (\texttt{c2}) for marginalized (\texttt{m}) and non-marginalized (\texttt{p}) agents are modeled as normal-distributed values with a \texttt{mean} and a \texttt{deviation} using sliders in the form of \texttt{<p|m>\_<c1|c2>\_noise\_<deviation|mean>}.

\begin{figure}[!htb]
    \centering
    \includegraphics[scale=.465]{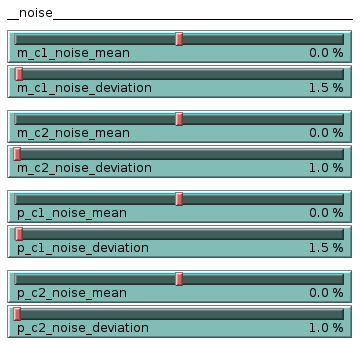}
    \caption{Screenshot of the NetLogo model's noise sliders.}
    \label{fig:noise}
\end{figure}

\section{Basic Validation}\label{text:validation}

This section provides some experimental results generated by the prototype. Note that this is not yet an attempt at full validation of the model and only serves to show how the model behaves. The two simulation runs with different conviction initialization values and otherwise identical configurations are compared to observe potential behavioral differences. The next subsection describes the configuration, followed by an outline of the results and noteworthy observations.

\subsection{Configuration}

The model is configured as follows; the full configuration is provided in Appendix \ref{text:config}. The \texttt{population} consists of 500 agents, with a \texttt{margin\_size} of 10.5\%. The \texttt{stealth} likelihood is negligibly low at 1\%, and \texttt{critical\_faculty} is set to 50\%. Ergo, marginalized agents are almost always read as such, and perpetrators of microaggressions accept criticism in half of the cases. A relatively low \texttt{positive\_threshold} has been chosen at 50\%, allowing agents to commit microaggressions even if they only slightly agree with \texttt{c1}. This is meant to model the unintentional nature of many microaggressions. The \texttt{action\_threshold} is only slightly higher at 66.6\%. Contrarily, the \texttt{negative\_threshold} is fairly low at 15\%, requiring high disagreement with \texttt{c1} for agents to react negatively. Non-marginalized agents are initialized to have generally higher agreements with both convictions, although there is significant variance. Marginalized agents generally have lower agreement with \texttt{c1}, and very low agreement with \texttt{c2}. For trial 1, \texttt{p\_c1\_mean} is set to 45\%; for trial 2, \texttt{p\_c1\_mean} is set to 66.6\%. This models a society where microaggressions are less and more common, respectively.

\subsection{Results}

\paragraph{Trial 1: \texttt{p\_c1\_mean = 45}}

\begin{figure}[!htb]
    \centering
    \includegraphics[width=.49\textwidth]{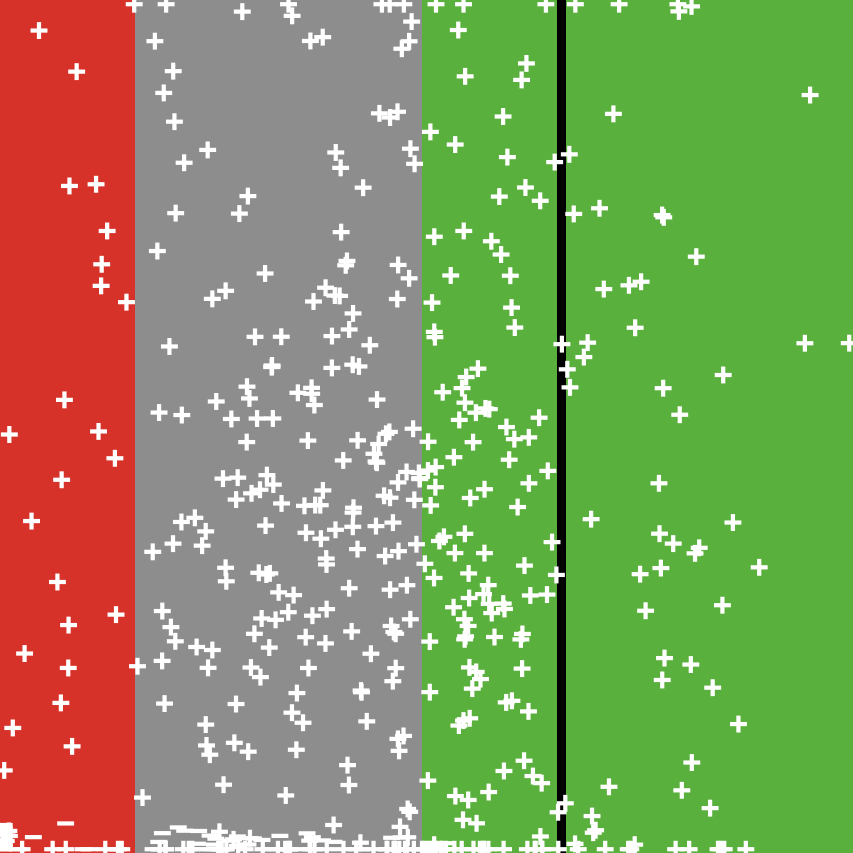}
    \includegraphics[width=.49\textwidth]{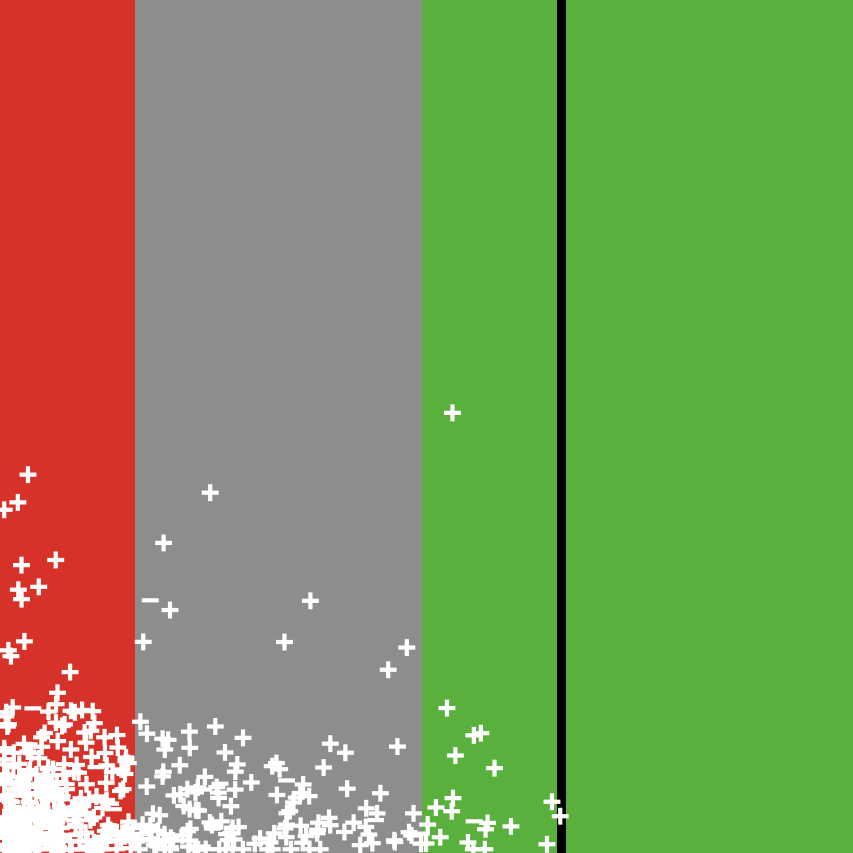}
    \\[\smallskipamount]
    \includegraphics[width=.49\textwidth]{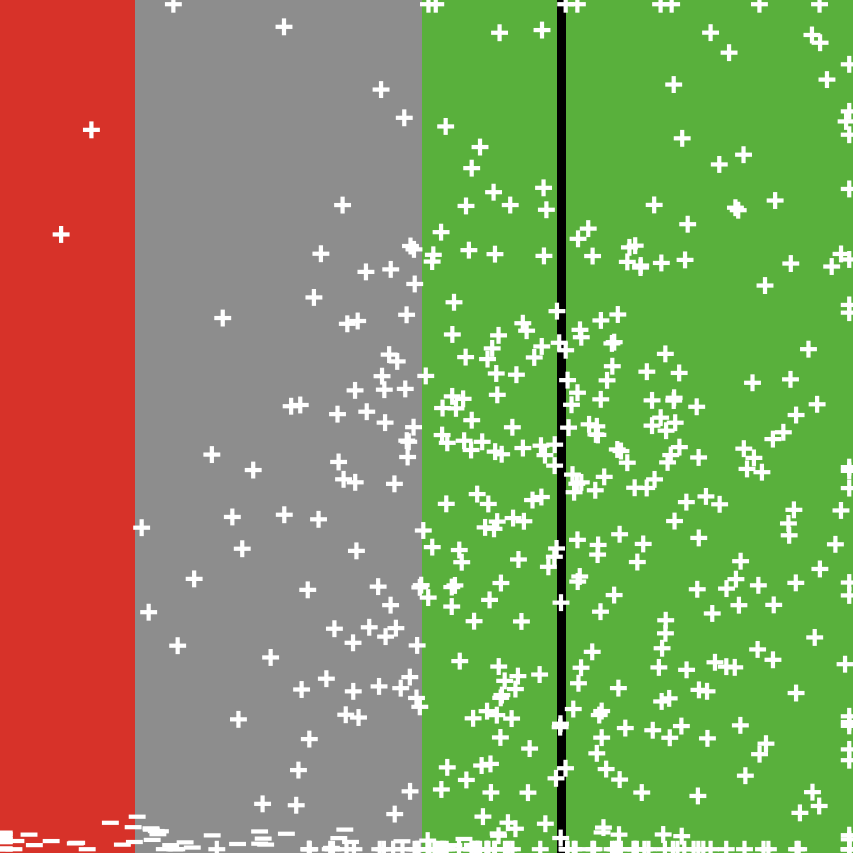}
    \includegraphics[width=.49\textwidth]{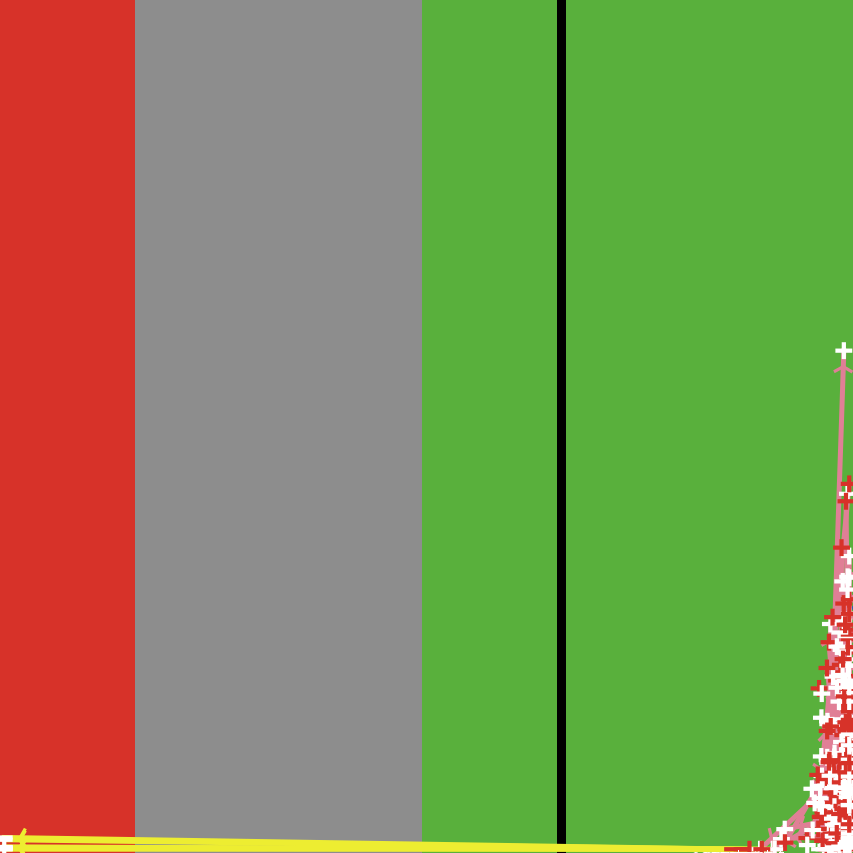}
    \caption{Visualization of the simulated society for trial 1 with \texttt{p\_c1\_mean = 45} on top and trial 2 with \texttt{p\_c1\_mean = 66.6} below at initialization on the left and after reaching the stop condition on the left.}
    \label{plot:trial}
\end{figure}

\begin{figure}[!htb]
    \centering
    \includegraphics[width=\textwidth]{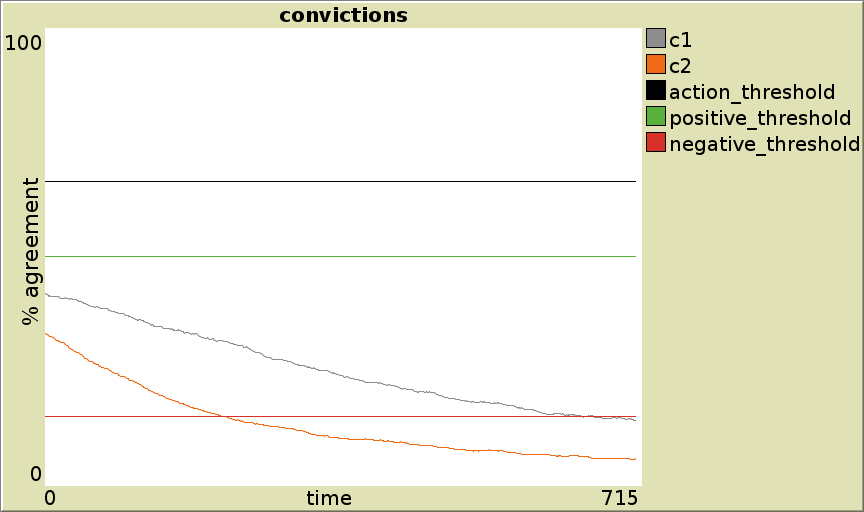}
    \\[\smallskipamount]
    \includegraphics[width=\textwidth]{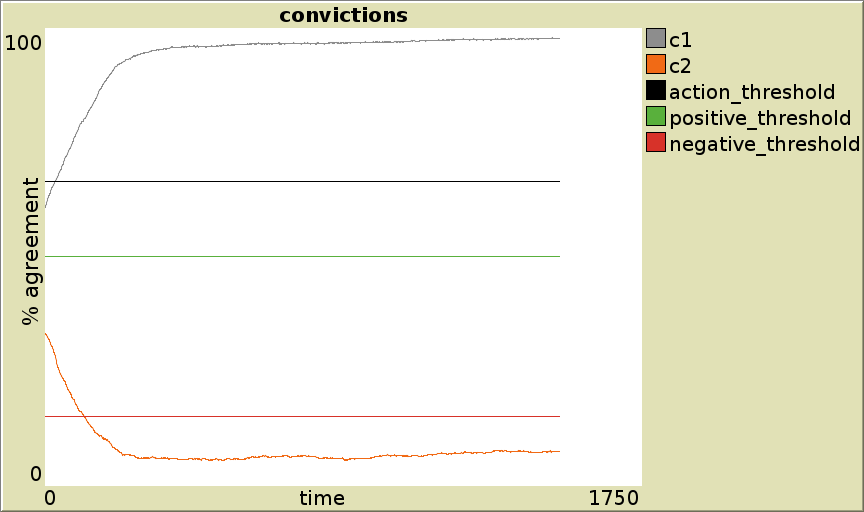}
    \caption{The convictions plot depicts the average agreement with \texttt{c1} (grey) and \texttt{c2} (orange) in the simulated society. For orientation, the \texttt{action\_threshold} (black), \texttt{positive\_threshold} (green), and \texttt{negative\_threshold} (red) are visualized as well. The upper plot represents trial 1, the lower plot represents trial 2.}
    \label{plot:convictions}
\end{figure}

\begin{figure}[!htb]
    \centering
    \includegraphics[width=\textwidth]{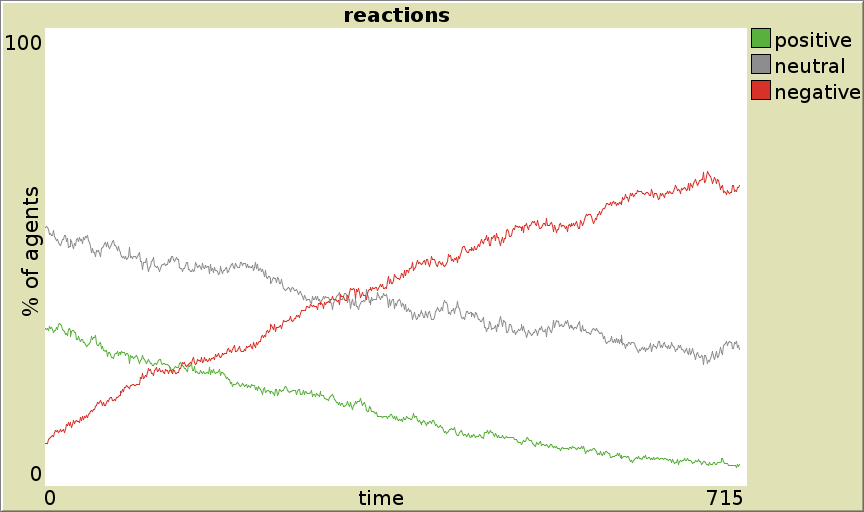}
    \\[\smallskipamount]
    \includegraphics[width=\textwidth]{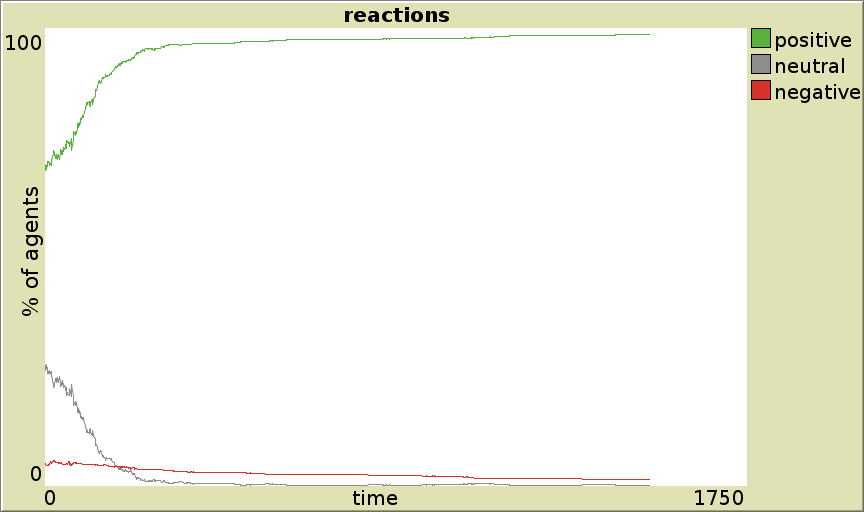}
    \caption{The reactions plot depicts the percentage of agents above the positive threshold (potential positive reactors, green), below the negative threshold (potential negative reactors, red), and in between the thresholds (neutral reactors, grey). The upper plot represents trial 1, the lower plot represents trial 2.}
    \label{plot:reactions}
\end{figure}

Trial 1 starts out with a fairly balanced \texttt{c1} and \texttt{c2} among non-marginalized agents, and balanced \texttt{c1} and low \texttt{c2} among marginalized agents (see top-left image in Figure \ref{plot:trial}). Both convictions show a mostly steady decline (see top image in \ref{plot:convictions}), resulting in a steadily growing number of negative reactors (see top image in Figure \ref{plot:reactions}). The simulation results in the stop condition \texttt{equilibrium: no potential perpetrators} after 710 ticks (see top-right image in Figure \ref{plot:trial}). At this point, overall average \texttt{c1} is just below the \texttt{negative\_threshold} of 15\%. 

\paragraph{Trial 2: \texttt{p\_c1\_mean = 66.6}}

Trial 2 starts out with generally higher \texttt{c1} and balanced \texttt{c2} among non-marginalized agents, and balanced \texttt{c1} and low \texttt{c2} among marginalized agents (see bottom-left image in Figure \ref{plot:trial}). At the start, about 80\% of non-marginalized agents are potential positive reactors. The value quickly rises and stalls at about 99\% (see top-right image in Figure \ref{plot:reactions_groups}). Marginalized agents start out at only about 9.5\% positive reactors, but follow this trend with a small delay and stalling at about 85\% (see bottom-right image in Figure \ref{plot:reactions_groups}). Overall average \texttt{c2} conviction falls and saturates at about 10\% (see bottom image in \ref{plot:convictions}). The simulation results in the stop condition \texttt{deadlock: society is too polarized for change} after 1508 ticks (see bottom-right image in Figure \ref{plot:trial}). At this point, overall average \texttt{c1} is about 97.5\% (see bottom image in Figure \ref{plot:convictions}) with the remaining agents maintaining low \texttt{c1} mostly belonging to the marginalized community (see left column in Figure \ref{plot:convictions_groups}).

\section{Discussion}

Intuitively, the behavior seen in the trials seems plausible. Lower initial \texttt{c1}, as seen in trial 1, allows the agents to slowly be convinced of the harmful nature of microaggressions, while the higher initial \texttt{c1} of trial 2 results in a critical mass of agents being so radicalized that the society ends up becoming too polarized for change. This is an acceptable result for a first prototype, but the behavior should be analyzed and refined in future research.

The impact of the \texttt{noise} parameter requires further examination. With the low \texttt{negative\_threshold} and \texttt{positive\_threshold} used in this paper, the potential negative reactors have a much higher chance of being pushed above \texttt{negative\_threshold} than a potential positive reactor has of being pulled below \texttt{positive\_threshold}. Implementing different noise parameters for agents in different threshold ranges would be possible.

With the configuration used in this paper (see Section \ref{text:validation}), overall \texttt{c2} still falls even in frequent perpetrators with high \texttt{c1}. The relevant parameters should be re-evaluated to allow more insights into the dynamic between \texttt{c1} and \texttt{c2}.

Due to NetLogo's simple design, complex constraints like intersectionality (see Section \ref{text:simulation}) can become challenging. Re-implementing the model in a simulation framework that supports an object-oriented class-based approach to dynamically model different types of marginalization might be advised.

\section{Conclusion and Outlook}

\subsection{Future Work}

The model should be improved and possibly re-implemented in a simulation framework that allows for more flexibility than NetLogo. This would make implementing additional configuration options for more granular control easier.

The model's default parameters should be refined and, if possible, validated against real-world empirical data. Different "scenarios" to explore the situation of specific real-world marginalized communities could be researched and pre-configured to be loaded in the model's user interface. Introducing a \texttt{false\_positive\_clock} configuration parameter analog to \texttt{stealth} could be added to model cases where agents might falsely read a non-marginalized agent as marginalized.

There are many possibilities for extending the background logic, such as allowing an agent's \texttt{c1} to influence their \texttt{critical\_faculty}, and introducing interdependency between an agent's \texttt{c2} and \texttt{c1}.

\subsection{Final Remarks}

This paper set out to apply a computational ethics approach to the phenomenon of microaggressions and the prejudices against marginalized communities that arise from it. A basic model has been theorized, and a prototype for a simulation has been implemented in NetLogo. The fundamental viability of the approach has been demonstrated, and possible expansions for future research have been discussed. Readers are encouraged to conduct their own experiments with the model, which has been made available under \url{https://neothethird.gitlab.io/ceth-seminar/model.nlogo}.

% \begin{itemize}
%   \item Modelling is hard. There are multiple ways to model the case one wants to inspect and choosing the right way is hard.
%   \item The complexity of analyzing the model grows with the complexity of the model itself
%   \item The results can only be seen in the presence of the model. If the model does not depict the reality, the conclusions and results of the model can not be transferred into reality.
% \end{itemize}

% \charactercount{main} chars

\bibliography{main}
% \pagebreak
\begin{appendices}
\section{Model Configuration} \label{text:config}

\begin{longtable}{>{\hspace{0pt}}m{0.45\linewidth}>{\hspace{0pt}}m{0.112\linewidth}}
\caption{General configuration.}\\
\texttt{population} & 500 \endfirsthead
\texttt{margin\_size} & 10.5 \\
\texttt{stealth} & 1 \\
\texttt{critical\_faculty} & 50
\end{longtable}

\begin{longtable}{>{\hspace{0pt}}m{0.45\linewidth}>{\hspace{0pt}}m{0.112\linewidth}}
\caption{Thresholds.}\\
\texttt{action\_threshold} & 66.6 \endfirsthead
\texttt{positive\_threshold} & 50 \\
\texttt{negative\_threshold} & 15
\end{longtable}

\begin{longtable}{>{\hspace{0pt}}m{0.45\linewidth}>{\hspace{0pt}}m{0.112\linewidth}}
\caption{Conviction initialization.}\\
\texttt{p\_c1\_mean} & 45 $\lor$ 66.6 \endfirsthead
\texttt{p\_c1\_deviation} & 20 \\
\texttt{p\_c2\_mean} & 33.3 \\
\texttt{p\_c2\_deviation} & 33.3 \\
\texttt{m\_c1\_mean} & 20 \\
\texttt{m\_c1\_deviation} & 20 \\
\texttt{m\_c2\_mean} & 1 \\
\texttt{m\_c2\_deviation} & 1 \\
\end{longtable}

\begin{longtable}{>{\hspace{0pt}}m{0.45\linewidth}>{\hspace{0pt}}m{0.112\linewidth}}
\caption{Conviction changes due to noise.}\\
\texttt{p\_c1\_noise\_mean} & 0 \endfirsthead
\texttt{p\_c1\_noise\_deviation} & 1.5 \\
\texttt{p\_c2\_noise\_mean} & 0 \\
\texttt{p\_c2\_noise\_deviation} & 1 \\
%  & \\
\texttt{m\_c1\_noise\_mean} & 0 \\
\texttt{m\_c1\_noise\_deviation} & 1.5 \\
\texttt{m\_c2\_noise\_mean} & 0 \\
\texttt{m\_c2\_noise\_deviation} & 1
\end{longtable}

\begin{longtable}{>{\hspace{0pt}}m{0.45\linewidth}>{\hspace{0pt}}m{0.112\linewidth}}
\caption{Conviction changes due to Idleness.}\\
\texttt{p\_c1\_on\_idle} & -0.1 \endfirsthead
\texttt{p\_c2\_on\_idle} & -0.1 \\
%  & \\
\texttt{m\_c1\_on\_idle} & -0.1 \\
\texttt{m\_c2\_on\_idle} & -0.1
\end{longtable}

\begin{longtable}{>{\hspace{0pt}}m{0.45\linewidth}>{\hspace{0pt}}m{0.112\linewidth}}
\caption{Conviction changes due to positive reactions.}\\
\texttt{p\_c1\_on\_positive\_to\_p} & 2.5 \endfirsthead
\texttt{p\_c1\_on\_positive\_from\_p} & 5 \\
\texttt{p\_c1\_on\_positive\_to\_m} & 2.5 \\
\texttt{p\_c1\_on\_positive\_from\_m} & 5 \\
%  & \\
\texttt{p\_c2\_on\_positive\_to\_p} & 0 \\
\texttt{p\_c2\_on\_positive\_from\_p} & 0 \\
\texttt{p\_c2\_on\_positive\_to\_m} & 0 \\
\texttt{p\_c2\_on\_positive\_from\_m} & 0 \\
%  & \\
\texttt{m\_c1\_on\_positive\_to\_p} & 2.5 \\
\texttt{m\_c1\_on\_positive\_from\_p} & 5 \\
\texttt{m\_c1\_on\_positive\_to\_m} & 2.5 \\
\texttt{m\_c1\_on\_positive\_from\_m} & 5 \\
%  & \\
\texttt{m\_c2\_on\_positive\_to\_p} & 0 \\
\texttt{m\_c2\_on\_positive\_from\_p} & 0 \\
\texttt{m\_c2\_on\_positive\_to\_m} & 0 \\
\texttt{m\_c2\_on\_positive\_from\_m} & 0
\end{longtable}

\begin{longtable}{>{\hspace{0pt}}m{0.45\linewidth}>{\hspace{0pt}}m{0.112\linewidth}}
\caption{Conviction changes due to neutral reactions.}\\
\texttt{p\_c1\_on\_neutral\_to\_p} & 1 \\
\texttt{p\_c1\_on\_neutral\_from\_p} & 2.5 \\
\texttt{p\_c1\_on\_neutral\_to\_m} & 2 \\
\texttt{p\_c1\_on\_neutral\_from\_m} & 2.5 \\
%  & \\
\texttt{p\_c2\_on\_neutral\_to\_p} & 0 \\
\texttt{p\_c2\_on\_neutral\_from\_p} & 0 \\
\texttt{p\_c2\_on\_neutral\_to\_m} & 0 \\
\texttt{p\_c2\_on\_neutral\_from\_m} & 0 \\
%  & \\
\texttt{m\_c1\_on\_neutral\_to\_p} & 1 \\
\texttt{m\_c1\_on\_neutral\_from\_p} & 2.5 \\
\texttt{m\_c1\_on\_neutral\_to\_m} & 2 \\
\texttt{m\_c1\_on\_neutral\_from\_m} & 2.5 \\
%  & \\
\texttt{m\_c2\_on\_neutral\_to\_p} & 0 \\
\texttt{m\_c2\_on\_neutral\_from\_p} & 0 \\
\texttt{m\_c2\_on\_neutral\_to\_m} & 0 \\
\texttt{m\_c2\_on\_neutral\_from\_m} & 0
\end{longtable}

\begin{longtable}{>{\hspace{0pt}}m{0.45\linewidth}>{\hspace{0pt}}m{0.112\linewidth}}
\caption{Conviction changes due to negative reactions.} \\
\texttt{p\_c1\_on\_negative\_to\_p} & -5 \\
\texttt{p\_c1\_on\_negative\_accepted\_from\_p} & -10 \\
\texttt{p\_c1\_on\_negative\_rejected\_from\_p} & 15 \\
%  & \\
\texttt{p\_c1\_on\_negative\_to\_m} & -10 \\
\texttt{p\_c1\_on\_negative\_accepted\_from\_m} & -10 \\
\texttt{p\_c1\_on\_negative\_rejected\_from\_m} & 30 \\
%  & \\
\texttt{p\_c2\_on\_negative\_to\_p} & -10 \\
\texttt{p\_c2\_on\_negative\_accepted\_from\_p} & -10 \\
\texttt{p\_c2\_on\_negative\_rejected\_from\_p} & 0 \\
%  & \\
\texttt{p\_c2\_on\_negative\_to\_m} & -50 \\
\texttt{p\_c2\_on\_negative\_accepted\_from\_m} & -50 \\
\texttt{p\_c2\_on\_negative\_rejected\_from\_m} & 50 \\
%  & \\
%  & \\
\texttt{m\_c1\_on\_negative\_to\_p} & -5 \\
\texttt{m\_c1\_on\_negative\_accepted\_from\_p} & -10 \\
\texttt{m\_c1\_on\_negative\_rejected\_from\_p} & 15 \\
%  & \\
\texttt{m\_c1\_on\_negative\_to\_m} & -10 \\
\texttt{m\_c1\_on\_negative\_accepted\_from\_m} & -10 \\
\texttt{m\_c1\_on\_negative\_rejected\_from\_m} & 30 \\
%  & \\
\texttt{m\_c2\_on\_negative\_to\_p} & -10 \\
\texttt{m\_c2\_on\_negative\_accepted\_from\_p} & -10 \\
\texttt{m\_c2\_on\_negative\_rejected\_from\_p} & 0 \\
%  & \\
\texttt{m\_c2\_on\_negative\_to\_m} & -50 \\
\texttt{m\_c2\_on\_negative\_accepted\_from\_m} & -50 \\
\texttt{m\_c2\_on\_negative\_rejected\_from\_m} & 50
\end{longtable}

\section{Plots}\label{text:plots}

\begin{figure}[H]
    \centering
    \includegraphics[width=\textwidth]{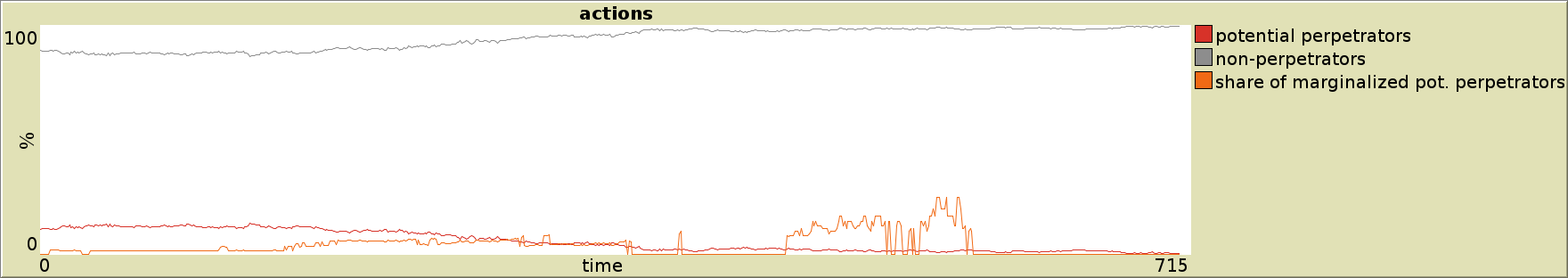}
    \\[\smallskipamount]
    \includegraphics[width=\textwidth]{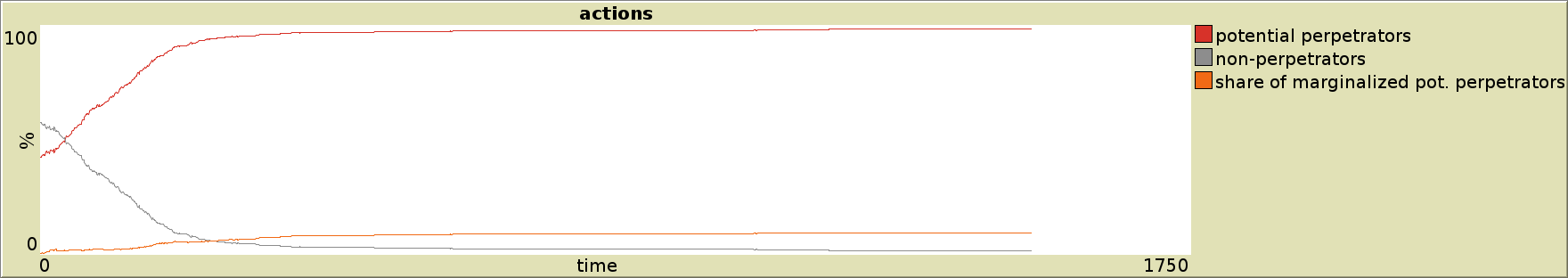}
    \caption{The actions plots depicts the percentage of potential perpetrators (red), the percentage of non-perpetrators (grey), and the share of marginalized agents among the potential perpetrators (orange) for trial 1 on top and trial 2 below.}
    \label{plot:actions}
\end{figure}

\begin{figure}[H]
    \centering
    \includegraphics[width=0.49\textwidth]{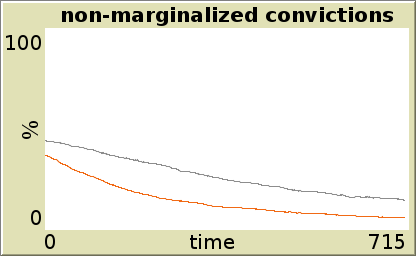}
    \includegraphics[width=0.49\textwidth]{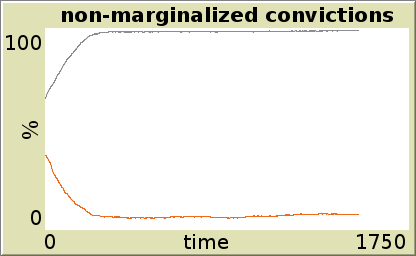}
    \\[\smallskipamount]
    \includegraphics[width=0.49\textwidth]{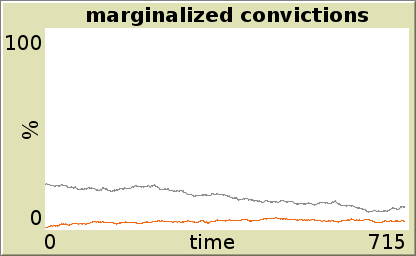}
    \includegraphics[width=0.49\textwidth]{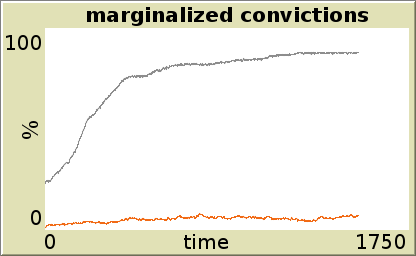}
    \caption{The (non-)marginalized convictions plots depict the average agreement with \texttt{c1} (grey) and \texttt{c2} (orange) among non-marginalized (top row) and marginalized (bottom row) agents in the simulated society for trial 1 on the left and trial 2 on the right.}
    \label{plot:convictions_groups}
\end{figure}

\begin{figure}[H]
    \centering
    \includegraphics[width=0.49\textwidth]{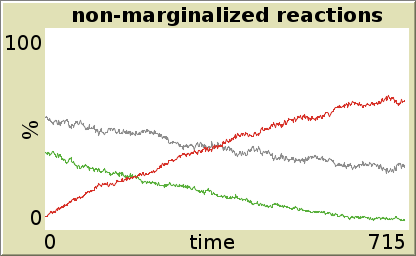}
    \includegraphics[width=0.49\textwidth]{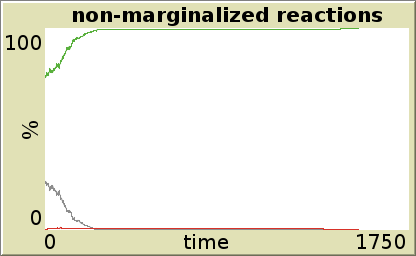}
    \\[\smallskipamount]
    \includegraphics[width=0.49\textwidth]{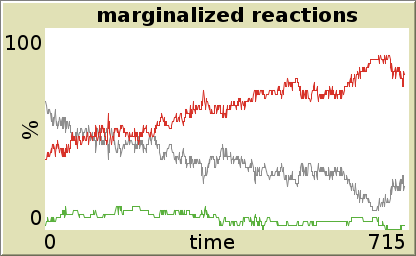}
    \includegraphics[width=0.49\textwidth]{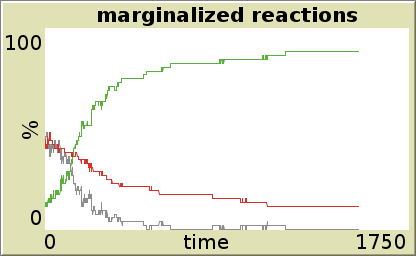}
    \caption{The (non-)marginalized reactions plots depict the percentage of non-marginalized (top row) and marginalized (bottom row) agents above the positive threshold (potential positive reactors, green), below the negative threshold (potential negative reactors, red), and in between the thresholds (neutral reactors, grey) for trial 1 on the left and trial 2 on the right.}
    \label{plot:reactions_groups}
\end{figure}

\end{appendices}
\end{document}